\documentclass[9pt,twocolumn,twoside]{osajnl}

\journal{journal} 

\setboolean{shortarticle}{true}

\title{The multiplexed light storage of Orbital Angular Momentum based on atomic ensembles}

\author[1]{Xin Yang}
\author[2]{Hong Chang}
\author[1]{Jinwen Wang}
\author[2]{Yan Ma}
\author[1]{Yun Chen}
\author[1]{Shuwei Qiu}
\author[1]{Zehao Shen}
\author[1]{Chengyuan Wang}
\author[1]{Quan Quan}
\author[1]{Dong Wei}
\author[1]{Haixia Chen}
\author[2]{Mingtao Cao}
\author[1,*]{Hong Gao}
\author[1]{Fuli Li}

\affil[1]{Ministry of Education Key Laboratory for Nonequilibrium Synthesis and Modulation of Condensed Matter, Shaanxi Province Key Laboratory
of Quantum Information and Quantum Optoelectronic Devices, School of Physics, Xi'an Jiaotong University, Xi'an, 710049, China}
\affil[2]{Key Laboratory of Time and Frequency Primary Standards, National Time Service Center, Chinese Academy of Sciences, Xi’an 710600, China}

\affil[*]{Corresponding author:honggao@xjtu.edu.cn}

\begin{abstract}
The improvement of the multi-mode capability of quantum memory can further improve the utilization efficiency of the quantum memory and reduce the requirement of quantum communication for storage units. In this letter, we experimentally investigate the multi-mode light multiplexing storage of orbital angular momentum (OAM) mode based on rubidium vapor, and demultiplexing by a photonic OAM mode splitter which combines a Sagnac loop with two dove prisms. Our results show a mode extinction ratio higher than 80$\%$ at 1 $\mu$s of storage time. Meanwhile, two OAM modes have been multiplexing stored and demultiplexed in our experimental configuration. We believe the experimental scheme may provide a possibility for high channel capacity and multi-mode quantum multiplexed quantum storage based on atomic ensembles. 
 
\end{abstract}

\setboolean{displaycopyright}{true}

\begin{document}

\maketitle

Qauntum memories, enabling the storage of an input photonic qubit and retrieval on controllable time, which can beat any classical device and constitute essential components in quantum repeaters and optical quantum information processing\cite{lvovsky2009optical,zhao2009long}. Over the past years, various memory protocols have been proposed, such as electromagnetically induced transparency (EIT)\cite{1998EIT}, gradient echo storage\cite{2012GEM}, optical frequency comb \cite{2019OFC}, Raman scheme \cite{2016raman} and the Duan–Lukin–Cirac–Zoller protocol \cite{2001DLCZ}. The EIT storage protocol is widely used among these storage protocols because of its simple configuration and easy implementation.

Improving the multi-mode storage capability in the spatial domain can effectively improve the working efficiency of quantum repeaters and reduce the requirement for quantum communication on the storage unit. Orbital angular momentum (OAM) beam is an attractive light field mode, and its topological charge can be taken as an arbitrary integer\cite{shen2019optical}. In principle, an infinite-dimensional Hilbert space can be constructed to realize higher dimensional spatial mode encoding of photon OAM. Therefore, using the OAM mode for storage is an effective way to realize efficient quantum repeaters. In recent years, research on the storage of photon OAM light modes has been carried out  \cite{nicolas2014quantum,yang2018multiplexed,wang2021efficient}. These studies mainly focus on storing low-order vector beams and finite-dimensional photon OAM quantum states \cite{parigi2015storage,ye2019experimental, wang2021efficient}, but the high-dimensional properties of OAM have not been fully utilized in any quantum memory experiment and multi-OAM mode storage simultaneously based on atomic ensembles has not been studied.

It is well known that effectively identifying the OAM mode is significant in multi-mode storage. To date, there are many traditional measurement methods for the photon OAM quantum state, including the round-hole diffraction measurement \cite{2008method}, optical transformation method \cite{2011measuring} and interference method \cite{2013phase,1996experiment}. However, the schemes of state identification of OAM quantum states after storage are all based on the projection measurement scheme\cite{2001entanglement}; The projection measurement scheme can only be used for the identification of the OAM quantum states but cannot achieve the separation of the OAM quantum states.An applicable quantum memory with multimode capacity should be achieved multiplexed storage and demultiplexed retrieval simultaneously. Therefore, if photons encoded with high-dimensional OAM in a quantum memory, not only the multimode storage shall be realized but also the stored OAM quantum states need to be efficiently separated after storage.  




\begin{figure*}[htbp]
\centering\includegraphics[width=0.95\linewidth]{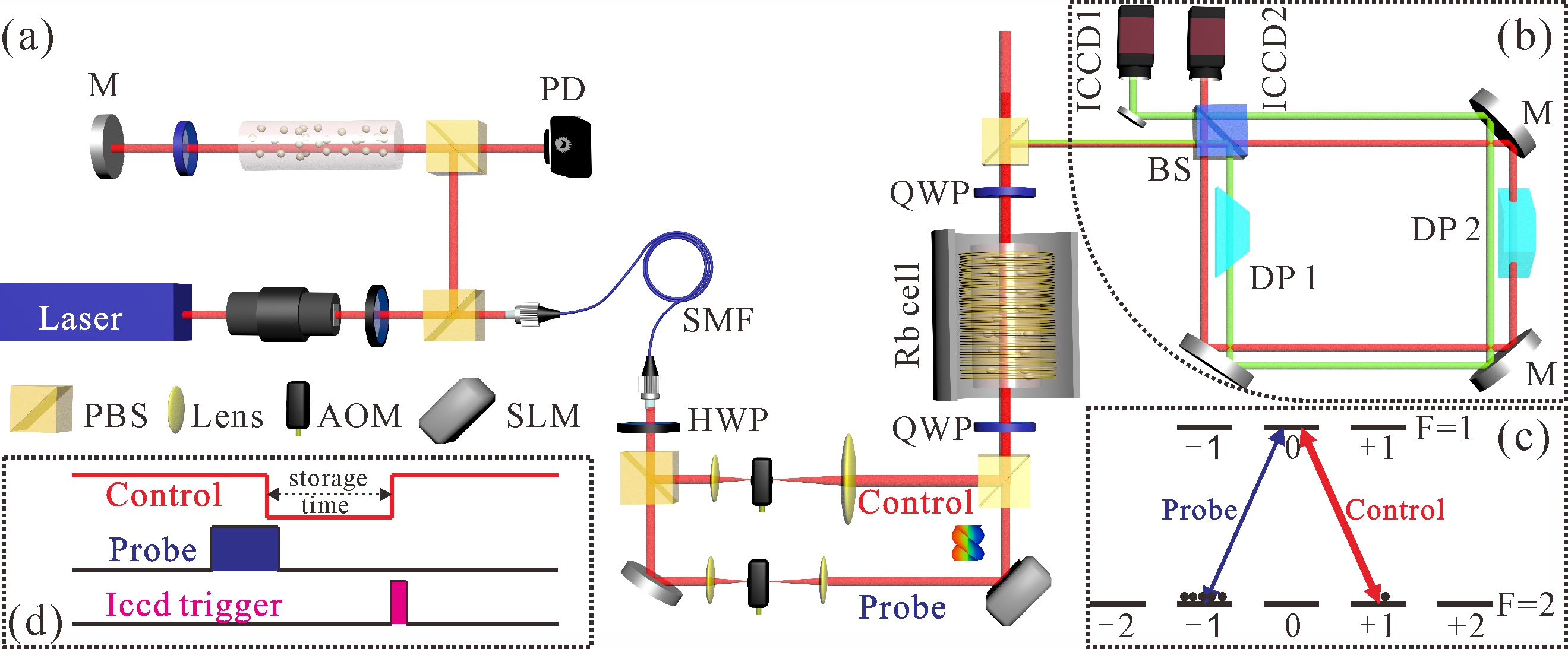}
\caption{(a)The experimental setup of Multiplexing quantum memory. PBS, polarization beam splitter; BS, beam splitter; HWP, half-wave plate; QWP, quarter-wave plate; SLM, spatial light modulator. (b) The setup of demultiplexing; DP, Dove prism; ICCD, intensified charge coupled device (c) Schematic energy diagram of $^{87}$Rb and light coupling.}
\label{fig1}
\end{figure*}

In this letter, we experimentally investigate multi-mode quantum multiplexing storage and demultiplexing based on rubidium vapor. Specifically, OAM beams with different topological charges have been stored in atomic ensembles through the use of electromagnetic induced transparency storage protocol and demultiplexed into different output ports by using the photon OAM mode splitter which combines a Sagnac loop with two dove prisms. Our results show a mode extinction ratio higher than 80 $\%$ at 1 $\mu$s of storage time. Meanwhile, two OAM modes have been multiplexed storage and demultiplexed output in different ports by using an OAM mode splitter. The results show that our experimental system can absolutely separate these two OAM modes after storage. We believe that the experimental scheme has applications in high channel capacity multi-mode quantum multiplexing storage and demultiplexing in atomic ensembles.

The experimental setup is shown in Fig.\ref{fig1}. The output of a 795 nm external cavity diode laser is divided into two parts. One part is applied for frequency locking, and the other one is sent through a single-mode fiber (SMF) to improve the spatial mode. After the fiber, the beam passes through a half-wave plate and polarization beam splitter to control the intensity of the control and probe beams. The transmitted part is chosen as the probe beam with the waist diameter of 2 mm (shown in Fig.\ref{fig1}(a)), while the reflected part is selected as the control beam, whose size is expanded three times larger than the probe beam by a telescope. The acousto-optic modulations(AOM) are inserted into these two parts for pulse modulation. The spatial light modulator(HAMAMATSU-X10468-07) is used for loading grating to generate the OAM light in the probe beam part. For effective imaging, the probe beam with OAM passed through the 4f imaging system (not shown in the Fig.\ref{fig1}) and propagates collinearly with the control beam. Finally, Both beams incident into the Rubidium cell and with the corresponding Zeeman sublevels of $^{87}$Rb. The energy-level configuration is shown in Fig.\ref{fig1}(c). In our experiment, the laser frequency is locked to the $5S_{1/2},F=2 \to 5P_{1/2},F'=1$ transition of the $^{87}$Rb D$_1$ line. The Rb cell has a length of 50 mm. A three-layer $\mu$ -metal magnetic shield isolates the cell from ambient magnetic fields. In our experiment, the power of the control and probe beam are The temperature of the cell is set to 65°C with a controller.

The control and the probe beams are separated into the reflected and transmitted parts using a QWP and a PBS after the Rb cell, and the reflected part is the probe beam carrying OAM. Then a Sagnac loop with two Dove prisms is applied to demultiplex OAM beams after storage. As shown in Fig.\ref{fig1}(b), the OAM probe beam is injected into the Sagnac loop and divided into two parts. These two parts go through the same path (clockwise and counterclockwise paths) and inject into the Dove prism with different rotation angles. The Dove prisms in the two paths produce different phases, and finally, these two parts interfere with each other in the beam splitter passes through the Sagnac loop. The spatial intensity distribution of the retrieval signal is recorded by an intensified charge-coupled device camera with a shutter time of 2 nanoseconds (ICCD,ANDOR TECHNOLOGY), as shown in Fig.\ref{fig1}(b).

Now let us focus on the experimental procedure shown in Fig.\ref{fig1}(d). First, the probe beam pulse with 5$\mu$s is input to the spatial light modulator(SLM) before the rubidium cell, and an OAM light pulse is generated to load a fork-shaped grating. In one-third of the back edge of the probe light pulse, the control light is turned off and then switched on again by using an AOM after a controllable time. The retrieved probe beam can be demultiplexed after the sagnac loop and detected by ICCD1 or ICCD2, depending on the topological charge of the probe light pulse. Secondly, our system realizes the multiplexing memory and demultiplexing of OAM states using the OAM mode splitter. To clearly understand the principle of demultiplexing based on the Sagnac loop, we assume that the quantum state of the probe beam is $ \left | l  \right \rangle$($l$ can be any integer). The transmitted and reflected states become $\frac{1}{2} \left | l  \right \rangle$ and $\frac{i}{2} \left | l  \right \rangle$. The outgoing light at the ICCD1 and the ICCD2 ports are expressed as
\begin{subequations} \label{equ_1-1}
\begin{align} 
    T_{ICCD1} = \frac{i}{2}\exp(i(2l\alpha + \pi)) \left | l  \right \rangle + \frac{i}{2}\exp(i \pi) \left | l  \right \rangle \\
	T_{ICCD2} = - \frac{1}{2}\exp(i(2l\alpha + \pi)) \left | l  \right \rangle + \frac{1}{2}\exp(i \pi) \left | l  \right \rangle
\end{align}
\end{subequations}
The first term of Eq.\ref{equ_1-1} is the quantum state that undergoes a counterclockwise path, and the second term describes the quantum state that undergoes a clockwise path. $\alpha$ is the rotation angle of dove prism. Actually, we set the angle of Dove Prism1 (DP1) to 90 degrees experimentally. Thus, the expression above becomes
\begin{subequations} \label{equ_1-2}
\begin{align} 
    T_{ICCD1} = -\frac{i}{2}\exp(il \pi) \left | l  \right \rangle - \frac{i}{2} \left | l  \right \rangle \\
	T_{ICCD2} = \frac{1}{2}\exp(il \pi)) \left | l  \right \rangle - \frac{1}{2} \left | l  \right \rangle
\end{align}
\end{subequations}
According to the Eq.\ref{equ_1-2}, when the topological charge of the incident probe beam is an odd number, we can obtain the $ T_{ICCD1}=0$ and $ T_{ICCD2}=\left | l  \right \rangle$, which means that the probe beam with quantum state $\left | l  \right \rangle$ constructive interference in the ICCD2 port and destructive interference at the ICCD1 port. The output is reversed when the topological charge is an even number.

\section{Results and analysis}
\begin{figure}[tp]
\centering\includegraphics[width=0.95\linewidth]{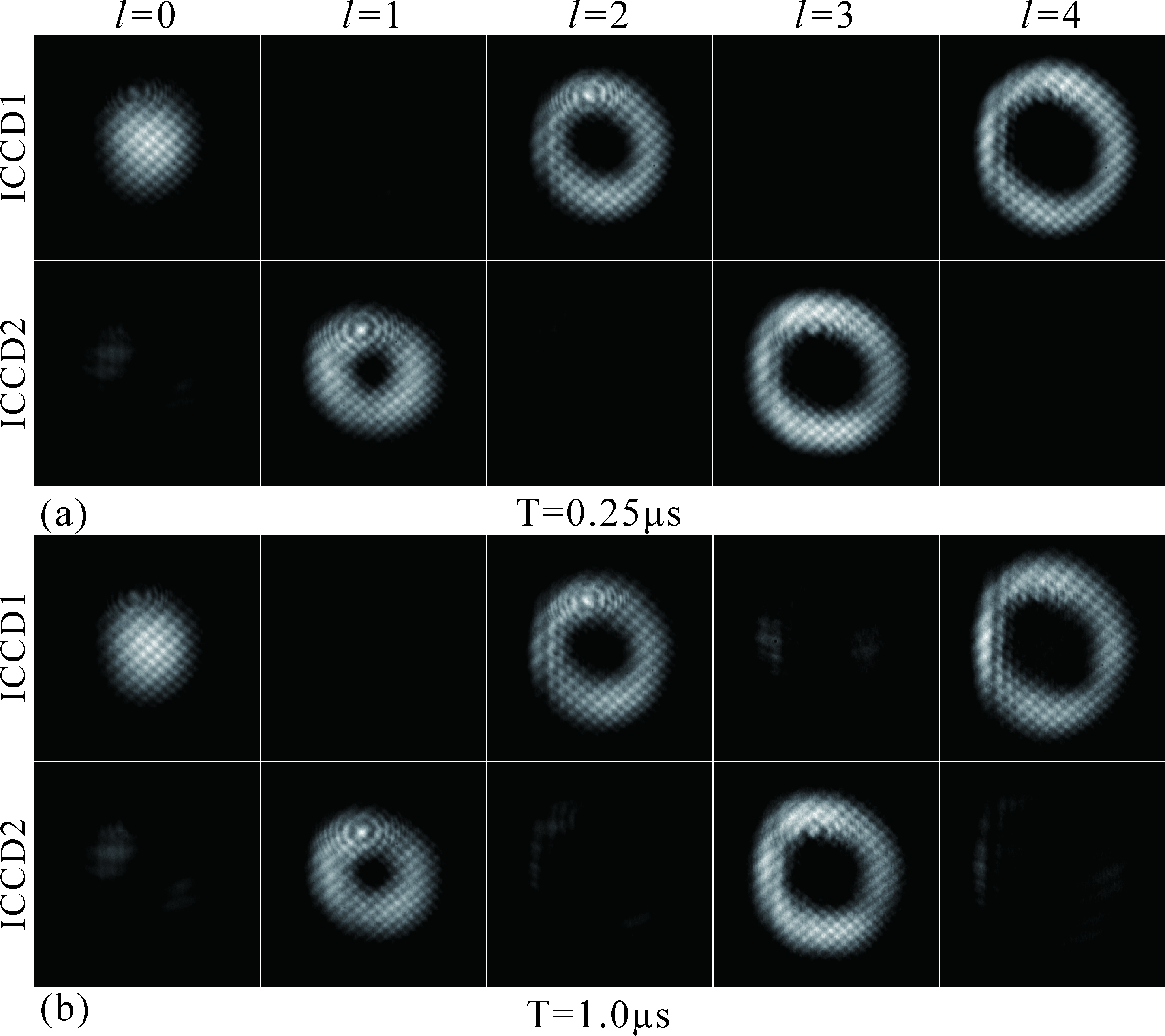}
\caption{(a) The demultiplexing of the probe beam after 0.25 $\mu$s of memory. (b) The demultiplexing of the probe beam after 1 $\mu$s of memory.}
\label{fig2}
\end{figure}

After checking the system reliability again, mainly including light leakage and ICCD normal trigger. We shown Our experimental results in Fig. \ref{fig2}(a). Clearly, the retrieval signal when carrying topological charge of 0, 2, 4 is captured by ICCD1, while, the signal carrying topological charge of 1 and 3 are captured by ICCD2 after 0.25$\mu$s storage time, respectively. The demultiplexing system still has good characteristics after 1 $\mu$s storage time as shown in Fig. \ref{fig2}(b). It can be seen that our demultiplexing system can separate the OAM mode very well and has a good mode contrast. In addition, we use oblique lens to realize self-interference of retrieval signal in experiment. The interference pattern is shown in the upper right corner of Fig. \ref{fig2}. As the results show that the OAM states remain unchanged after storage, and different OAM states are separated in different output channels.

\begin{figure}[htbp]
\centering\includegraphics[width=0.95\linewidth]{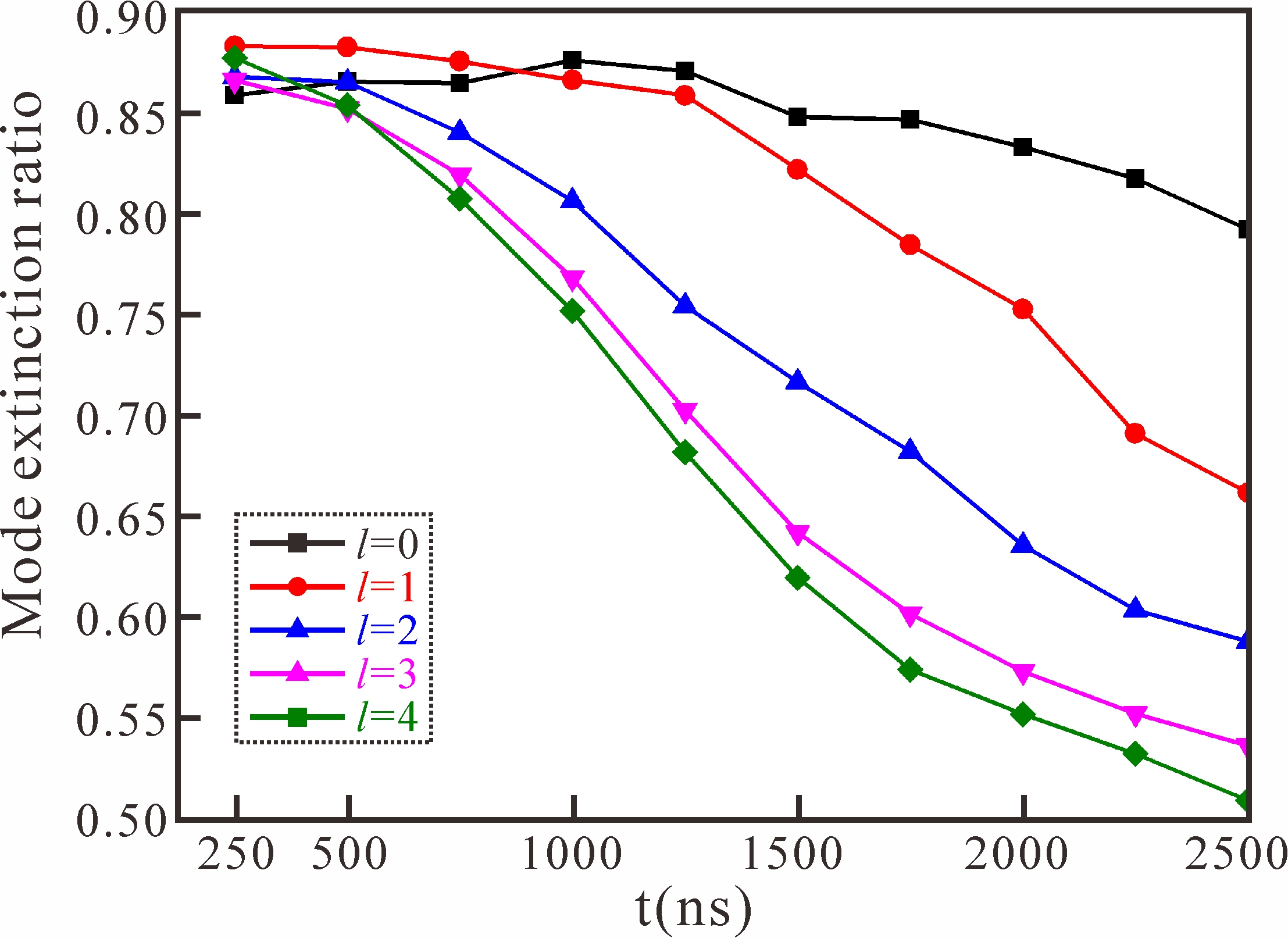}
\caption{The separation contrast of different OAM probe beams at different storage times. The black, red, blue, pink and green lines express the OAM probe beam with $l=0$, $l=1$, $l=2$, $l=3$ and $l=4$, respectively.}
\label{fig3}
\end{figure}

In order to study the quantum multiplexing memory and demultiplexing characteristics of OAM probe beams with different topological charges at different storage times. For effectively compare the retrieval mode quality under specific storage time, the mode extinction ratio is defined as  \begin{equation} \label{equ_1-4}
	V_{T_{ICCD1}/T_{ICCD2}} = |\frac{T_{ICCD1}-T_{ICCD2}}{ T_{ICCD1} + T_{ICCD2}}|
\end{equation}
We experimentally changed the clock switch of the control beam and the grating, and all results are shown in Fig. \ref{fig3}. 

We can obtain several differences by comparing the mode extinction ratio obtained from our demultiplexing system. It is observed that the Gauss pulse ($l=0$) maintains high mode extinction ratio about 80 $\%$ after 2.5 $\mu$s of quantum memory, as the black line shows in Fig. \ref{fig3}. The OAM pulse with $l=1$ remains an extinction ratio higher than 80 $\%$ after 1.5 $\mu$s of storage, as shown in the red line. However, the optical mode extinction ratio drops dramatically when the storage time exceeds 1.5 $\mu$s. The reason is due to the motion of the atomic spin wave that intensifies the diffusion of the readout optical mode when the optical mode carrying OAM is recorded into the atomic spin wave, which leads to a decrease in the extinction ratio after demultiplexing in our system. This influence becomes more pronounced when the optical mode carries higher-order topological charges, as shown in the blue line, pink and green lines of Fig. \ref{fig3}. 

\begin{figure}[htbp]
\centering\includegraphics[width=0.95\linewidth]{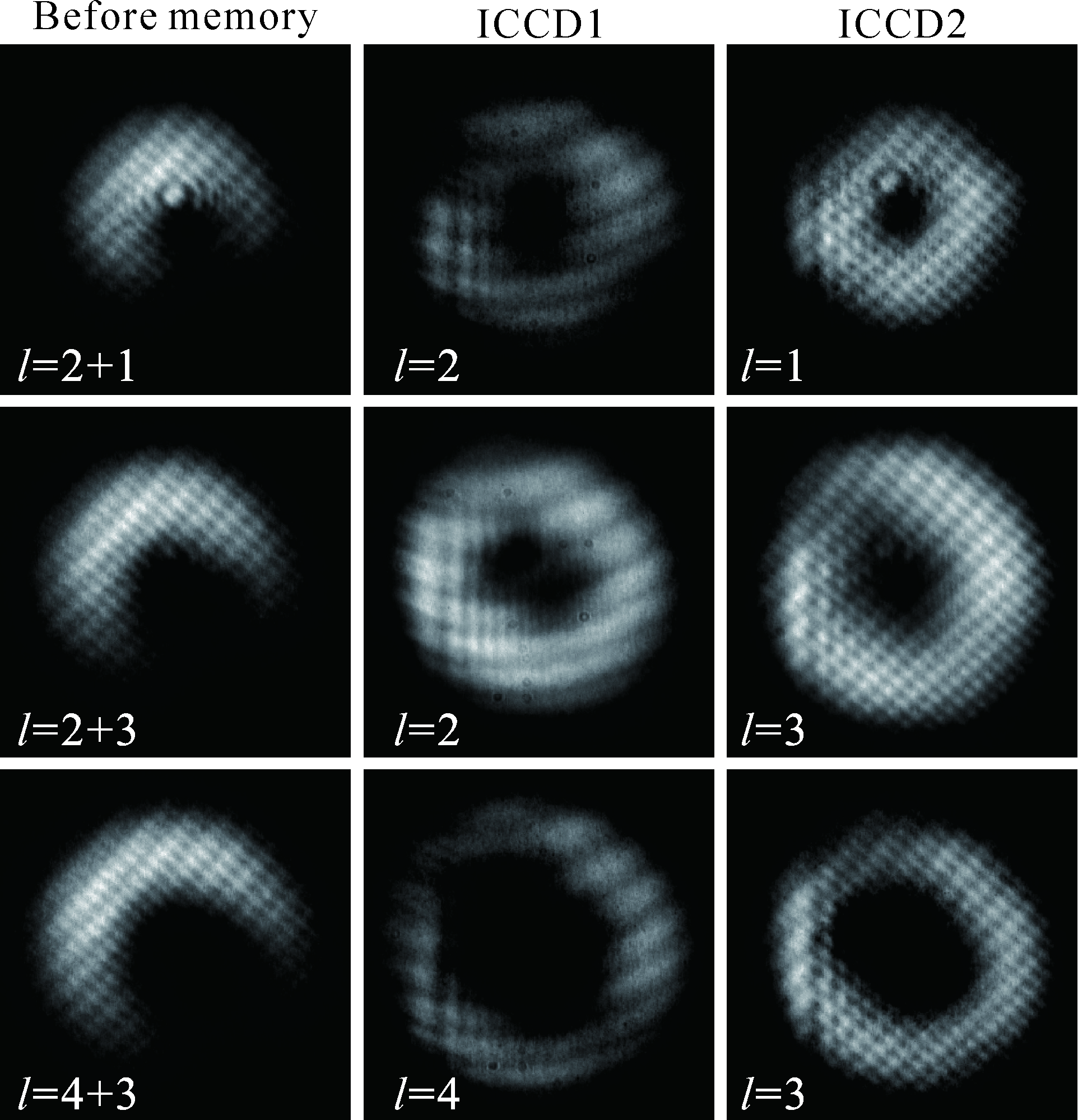}
\caption{Multiplexing memory and demultiplexing output of two OAM modes at 0.25 $\mu$s.}
\label{fig4}
\end{figure}

In order to verify the multi-mode capabilities of multiplexing storage and demultiplexing in our experimental setup. Two OAM modes expressed as $\left | 1  \right \rangle + \left | 2  \right \rangle$, $\left | 2  \right \rangle + \left | 3  \right \rangle$ and $\left | 3  \right \rangle + \left | 4  \right \rangle$ have been generated for multi-mode storage, and the intensity distribution as shown in the left column of Fig. \ref{fig4}. The figure shows that, two OAM modes injected into the memory system interfere to form a moon-like intensity distribution. After 0.25 $\mu$s stored, this two OAM modes with different topological charges are separated into different paths, as shown in the two right columns of Fig. \ref{fig4}. All odd OAM modes have been recorded by ICCD1, while ICCD2 has recorded the even OAM modes, and the separated OAM modes have a good intensity distribution. An effective OAM mode separator can not only effectively separate the OAM modes in different channels, but also maintain the OAM modes characteristic. Therefore, oblique lenses are applied again to achieve OAM mode self-interference. The experiment results are shown in the upper right corner of Fig. \ref{fig4}, the two OAM states remain unchanged after demultiplexing. At this point, our experimental system can not only realize the storage of the single OAM state and separated it in different ports by using an OAM modes splitter but can also store and separate two OAM modes simultaneously.

In conclusion, we have demonstrated the experimental realization of multi-mode multiplexed storage and demultiplexed of OAM beams at room temperature based on atomic ensembles. With the Sagnac interferometer embedding two dove prisms, we obtained a retrieval signal of the OAM mode with an odder or even topological charge in different interfering ports and a mode extinction ratio higher than 80 $\%$ at the specific storage time. Meanwhile, two OAM superimposed modes have been stored and demultiplexed successfully, and the results show that our experimental scheme can effectively separate the two OAM modes after quantum memory. The OAM demultiplexing devices based on the Sagnac loop structure not only have higher separation contrast but can also be extended to a higher dimensional multimode multiplexing storage and demultiplexing of OAM modes. It is worth noting that quantum memory with  multimode capacity not only needs to maintain the storage modes, but also needs to maintain the high fidelity. Our results may have applications in high channel capacity multi-mode quantum multiplexing storage and demultiplexing based on atomic ensembles.

\begin{backmatter}

\bmsection{Funding} 
National Natural Science Foundation of China (NSFC) (11374238, 11534008, 11574247, 11604257, 11604258, 11774286); National Science Foundation (NSF) (1602755); 

\bmsection{Disclosures} 

\smallskip


\bmsection{Disclosures} The authors declare no conflicts of interest.

\bmsection{Data Availability Statement} Data underlying the results presented in this paper are not publicly available at this time but may be obtained from the authors upon reasonable request.


\end{backmatter}


\bibliography{sample}

\bibliographyfullrefs{sample}



\ifthenelse{\equal{\journalref}{aop}}{%
\section*{Author Biographies}
\begingroup
\setlength\intextsep{0pt}
\begin{minipage}[t][6.3cm][t]{1.0\textwidth} 
  \begin{wrapfigure}{L}{0.25\textwidth}
    \includegraphics[width=0.25\textwidth]{john_smith.eps}
  \end{wrapfigure}
  \noindent
  {\bfseries John Smith} received his BSc (Mathematics) in 2000 from The University of Maryland. His research interests include lasers and optics.
\end{minipage}
\begin{minipage}{1.0\textwidth}
  \begin{wrapfigure}{L}{0.25\textwidth}
    \includegraphics[width=0.25\textwidth]{alice_smith.eps}
  \end{wrapfigure}
  \noindent
  {\bfseries Alice Smith} also received her BSc (Mathematics) in 2000 from The University of Maryland. Her research interests also include lasers and optics.
\end{minipage}
\endgroup
}{}

\end{document}